%% file: ms.tex
\documentclass[iop, numberedappendix]{emulateapj}		
\usepackage{apjfonts}

\newcommand{\hi}{H\,{\sc i}\rm}
\newcommand{\hii}{H\,{\sc ii}\rm}
\newcommand{\hiiit}{\mbox{H\,{\footnotesize II}}}

\newcommand{\heii}{He\,{\sc ii}\rm}

\newcommand{\nii}{[N\,{\sc ii}]}

\newcommand{\oiii}{[O\,{\sc iii}]}
\newcommand{\oii}{[O\,{\sc ii}]}
\newcommand{\oi}{[O\,{\sc i}]}

\newcommand{\sii}{[S\,{\sc ii}]}

\newcommand{\neiii}{[Ne\,{\sc iii}]}

\newcommand{\op}{O$^{+}$}

\newcommand{\opp}{O$^{++}$}

\newcommand{\expone}{$^{-1}$}

\newcommand{\gt}{\,$>$\,}
\newcommand{\lt}{\,$<$\,}

\newcommand{\nethree}{Ne$_{3868}$}
\newcommand{\othree}{O$_{5007}$}

\newcommand{\rs}{{\sc RS08}}

\newcommand{\eq}{\,=\,}

\newcommand{\te}{$T_e$}

\newcommand{\hgamma}{H$\gamma$}
\newcommand{\hdelta}{H$\delta$}
\newcommand{\hbeta}{H$\beta$}
\newcommand{\halpha}{H$\alpha$}
\newcommand{\lin}{$\,\lambda$}
\newcommand{\llin}{$\,\lambda\lambda$}

\newcommand{\oh}{12\,+\,log(O/H)}

\newcommand{\rtwothree}{R$_{23}$}
\newcommand{\vs}{vs.}

\slugcomment{}

\shorttitle{The abundance scatter in M33}
\shortauthors{F. Bresolin}

\begin{document}

\title{The abundance scatter in M33 from \hii\ regions:\\Is there any evidence for azimuthal metallicity variations?\footnotemark[1]\\[3mm]} 

\footnotetext[1]{Based on observations obtained at the Gemini Observatory, which is operated by the 
Association of Universities for Research in Astronomy, Inc., under a cooperative agreement 
with the NSF on behalf of the Gemini partnership: the National Science Foundation (United 
States), the Science and Technology Facilities Council (United Kingdom), the 
National Research Council (Canada), CONICYT (Chile), the Australian Research Council (Australia), 
Minist\'{e}rio da Ci\^{e}ncia e Tecnologia (Brazil) 
and Ministerio de Ciencia, Tecnolog\'{i}a e Innovaci\'{o}n Productiva (Argentina). }

\author{Fabio Bresolin} \affil{Institute for Astronomy, 2680 Woodlawn Drive, Honolulu, HI 96822, USA\\ }

\begin{abstract}
Optical spectra of 25 \hii\ regions in the inner two kpc of the M33 disk have been obtained with the GMOS spectrograph 
at the Gemini North telescope. The oxygen abundance gradient measured from the detection of the \oiii\lin4363 auroral line displays a scatter
of approximately 0.06 dex, a much smaller value than recently reported by Rosolowsky \& Simon in this galaxy. The analysis of the abundances 
for  a large sample of \hii\ regions derived from the \rtwothree\ strong-line indicator confirms that the scatter is small over the full disk of M33, consistent with the measuring 
uncertainties,
and comparable to what is observed in other  spiral galaxies. No evidence is therefore found for significant azimuthal variations in the present-day metallicity of the interstellar medium in this galaxy on spatial scales from $\sim 100$~pc to a few kpc.
A considerable fraction of M33  \hii\ regions with auroral line detections  show spectral features revealing sources of hard ionizing radiation (such as \heii\ emission and large \neiii, \oiii\ line fluxes). Since \rtwothree\ is shown to severely underestimate the oxygen abundances in such cases, care must be taken in chemical abundance studies of extragalactic \hii\ regions  based on this strong-line indicator.

\end{abstract}

 
\section{Introduction}
The occurrence of radial metallicity gradients in spiral galaxies is presently understood within the inside-out formation paradigm  as due to disk growth by means of radially-dependent  gas infall and star formation rate (\citealt{Matteucci:1989,Boissier:1999}).
Recent age estimates of  resolved stellar populations   as a function of galactocentric distance in nearby galaxies has provided 
convincing evidence for the inside-out evolution of disk galaxies (\citealt{Williams:2009, Gogarten:2010}).
At the same time, the study of present-day metallicity gradients in galaxies from \hii\ region spectroscopy (\citealt{Vila-Costas:1992, Zaritsky:1994, Kennicutt:2003}) continues to provide important  empirical constraints  for the complex set of parameters that describe galactic chemical evolution models (\citealt{Yin:2009,Fu:2009}).

Azimuthal variations in the chemical composition of external spiral galaxies are observationally much less constrained than radial gradients.
The mixing of the interstellar medium (ISM) occurs over a range of timescales, that grow with the extent of the spatial scale considered.
On scales between 0.1 and 1~kpc the effects of cloud collisions and expanding supershells, powered by stellar winds and supernova explosions, and combined with galactic differential rotation, lead to heavy element dispersal and mixing of the ISM on relatively short timescales, around $10^8$ yr (\citealt{Roy:1995}), which helps to explain  the high degree of chemical homogeneity observed 
in the ISM of galaxies (\citealt{Scalo:2004, Edmunds:2005}). In the case of the spiral galaxy M101, \citet{Kennicutt:1996} concluded that the local scatter in metallicity is significantly smaller than the  0.1-0.15~dex dispersion derived from the oxygen abundances of 41 \hii\ regions, a conclusion that is in agreement with the expected level of inhomogeneity of the ISM from hierarchical models of star formation ($\sim$0.05 dex, \citealt{Elmegreen:1998}).

More recently, \citet{Bresolin:2009a} measured the oxygen abundances of 28 \hii\ regions in NGC~300 using the \oiii\lin4363 auroral line as a temperature diagnostic, and found a scatter in the radial gradient of only 0.05~dex. 
On the other hand, in  M33
\citet{Rosolowsky:2008} found a substantial intrinsic fluctuation in the \hii\ region oxygen abundance, $\sim0.11$ dex, in addition to the scatter that is simply due to observational errors. In particular, in the central two kpc of the M33 disk the dispersion in O/H is 0.2~dex, with a peak-to-peak variation of 0.8~dex over a spatial scale of about 1~kpc.  This result appears difficult to reconcile with the short mixing timescale of the ISM and with the small scatter measured in the present-day metallicities of other nearby spirals.

In this paper a new sample of \hii\ regions located in the inner two kpc of M33 is presented, with the goal of verifying 
the substantial  local dispersion in the oxygen abundance detected by \citet[\eq \rs]{Rosolowsky:2008}. Measuring accurate \hii\ region abundances in the central regions of spiral galaxies is critical for radial gradient studies, but it is generally made difficult by the enhanced cooling resulting from the increased metal line emission, which hinders the detection of the weak electron temperature diagnostic lines, such as \oiii\lin4363 
 ({\citealt{Bresolin:2008}).
The observations
presented here alleviate this difficulty, thanks to high-quality detections of the \oiii\lin4363 line in \hii\ regions  with galactocentric distances as small as 0.2~kpc (Sect.~2). The chemical abundance analysis  is carried out 
complementing the new data with existing samples of \hii\ regions, using both the auroral line information and the \rtwothree\ strong-line abundance indicator (Sect~3). The systematic effect on \rtwothree-based abundances  induced by the hard ionizing radiation field detected in several of the M33 \hii\ regions  is discussed in Sec.~4.

\section{Observations}
Spectra of \hii\ regions in M33 were obtained with the Gemini Multi-Object Spectrograph (GMOS, \citealt{Hook:2004}) at the Gemini North facility. The targets were selected from  narrow-band  \halpha\ images of two 5\farcm5$\times$5\farcm5 GMOS fields in the inner parts of the  galaxy, centered approximately 2\farcm1 (0.83~kpc deprojected distance) W and 5\farcm4 (1.67~kpc)  S of the center.
The spectroscopic data were acquired in queue mode on October 17, 2009, using two multi-object masks, one per field, with 1\farcs2-wide slits.
The seeing conditions were $\sim$0\farcs7 during the observations of the W field, and $\sim$1\farcs2 for the S field.
In order to minimize the effects of the differential atmospheric refraction the data were acquired at airmasses smaller than 1.16.
Three 1800\,s exposures were secured for each of the two fields using the B600 grating, which provided spectra covering 
the 3500-5100\,\AA\ wavelength range at a spectral resolution of $\sim$5.5\,\AA. For some of the targets, depending on their spatial distribution, the spectral coverage extended up to $\sim$6000\,\AA.

{\sc iraf}\footnote{{\sc iraf} is distributed by the National Optical Astronomy Observatories, which are operated by the Association of Universities for Research in Astronomy, Inc., under cooperative agreement with the National Science Foundation.}
routines in the {\tt\small gemini/gmos} package were used  for electronic bias subtraction, flat field correction and wavelength calibration of the raw data frames. Observations of the spectrophotometric standard star Wolf 1346 yielded the flux calibration.
The final version of the spectra was obtained by averaging the three spectra corresponding to each individual slit.
The \hii\ region sample comprises 25 objects, whose locations are shown in Fig.~\ref{fig:image}. The celestial coordinates of the targets are summarized in Table~\ref{table:sample} (where objects are listed in order of decreasing declination), together with their galactocentric distances in kpc, deprojected adopting an inclination angle for M33 of 56 degrees (\citealt{Zaritsky:1989}), a position angle of the major axis of 23 degrees (\citealt{de-Vaucouleurs:1991}), and the center coordinates measured by  \citet{Massey:1996}. 
A distance of 840~kpc from \citet{Freedman:2001} was assumed.
The identification of the subcomponents of the \hii\ regions  given in column (5) of Table~\ref{table:sample} was taken from \citet{Hodge:2002}.

\begin{figure*}
\medskip
\center \includegraphics[width=0.75\textwidth]{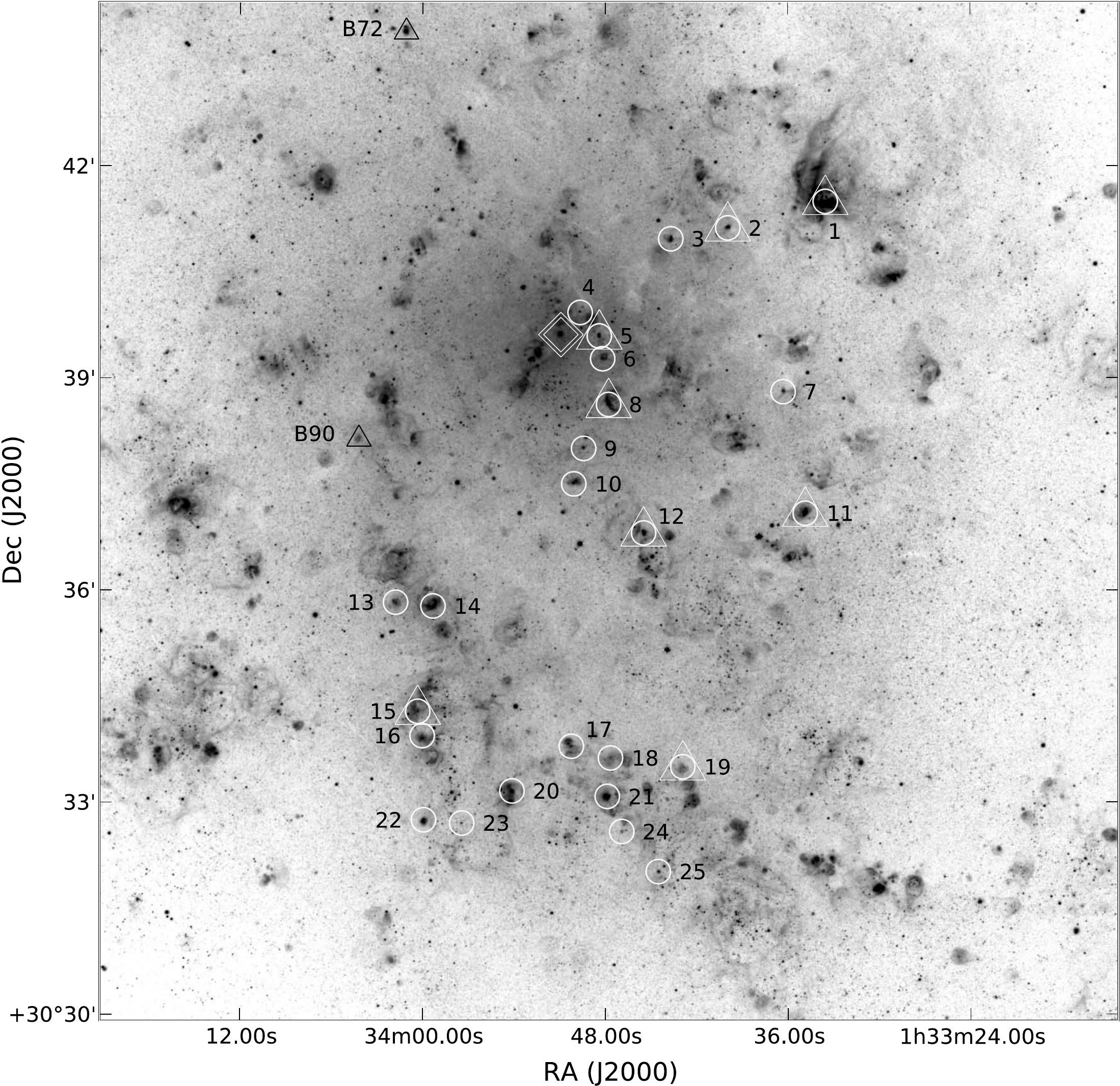}\medskip
\caption{Location of the \hii\/ regions studied in this work on a narrow-band \halpha\/ image of M33 (taken from the Local Group Galaxies Survey, \citealt{Massey:2006}). Targets where the \oiii\lin4363 line was detected are marked with triangles. The galaxy center is shown by the diamond symbol. The objects B\,72 and B\,90 were observed with the Subaru telescope by \citet{Bresolin:2010}. \label{fig:image}}
\end{figure*}

\input{tab1}

\subsection{Line intensities and oxygen abundances}
The emission  line intensities were measured with the {\tt\small splot} program in {\sc iraf} by integrating the fluxes under the line profiles.
The line intensities were corrected for interstellar reddening by assuming a case B intrinsic \hgamma/\hbeta\ ratio of 0.47 at \te\eq$10^4$\,K and the \citet{Seaton:1979} reddening law. 
The effects of the underlying stellar populations on the strength of the Balmer lines were accounted for by requiring that the \hgamma/\hbeta\ and \hdelta/\hbeta\ line ratios provided the same
value for the extinction. A median equivalent width of 2\,\AA\ was found for the absorption component, and was applied to the whole sample.
The resulting reddening-corrected emission line fluxes, normalized to \hbeta\,=\,100, are presented in Table~\ref{table:fluxes}.
The errors  account for uncertainties in the flat fielding, the flux calibration, the level of the continuum in proximity of each line and the logarithmic extinction coefficient $c$(\hbeta).

\input{tab2}

The electron temperature (\te) diagnostic line \oiii\lin4363  was measured for eight of the 25 \hii\ regions. The {\tt\small nebular} package in {\sc iraf} was used to derive \te\ for the \opp-emitting region, adopting an electron density of 10$^2$\,cm$^{-3}$ and the same atomic parameters used by \citet{Bresolin:2009a}.  The temperature in the \op-emitting region was estimated from the relation T\oii\eq 0.7\,T\oiii\ + 3000~K (\citealt{Garnett:1992}).
The \opp\ temperatures and the computed O/H abundance ratios are presented in columns (10) and (11) of Table~\ref{table:fluxes}, respectively.

\medskip
Ten objects in Table~\ref{table:sample} are in common with the work by RS08 (these are identified by the $*$ symbol in column 5). Moreover, NGC~595 (object 1 in Table~\ref{table:sample}) has also been recently observed with high resolution spectroscopy by \citet{Esteban:2009}. A  comparison with the line fluxes contained in these two publications is presented in Fig.~\ref{fig:comparison}. It can be seen that the agreement is generally acceptable, except for target 15 (C\,1Ab), for which RS08 measured a much stronger \oiii\lin5007 flux than the value reported here. It appears likely that the slits used by RS08 and in this work covered different portions of this high-excitation nebula. Fig.~\ref{fig:comparison}c indicates that \rtwothree, which can be used as a metallicity indicator\footnote{\rtwothree \eq (\oii\lin3727 + \oiii\llin4959,\,5007) / H$\beta$ (\citealt{Pagel:1979}).}, is generally consistent between the different works. The \oiii\lin4363-derived oxygen abundances  in Fig.~\ref{fig:comparison}d can differ between this work and \rs\ by significant amounts, up to 0.4 dex.  The logarithmic extinction  c(\hbeta) was found to be systematically higher by an average of 0.19 dex  in comparison with the RS08 values, for objects in common.
The comparison with the fluxes and the oxygen abundance published for NGC~595 by \citet[full symbols in Fig.~\ref{fig:comparison}]{Esteban:2009} is excellent. The c(\hbeta) value determined for this target is also in good agreement with their work (0.46 \vs\ 0.39).

\begin{figure}
\medskip
\epsscale{1.15}
\plotone{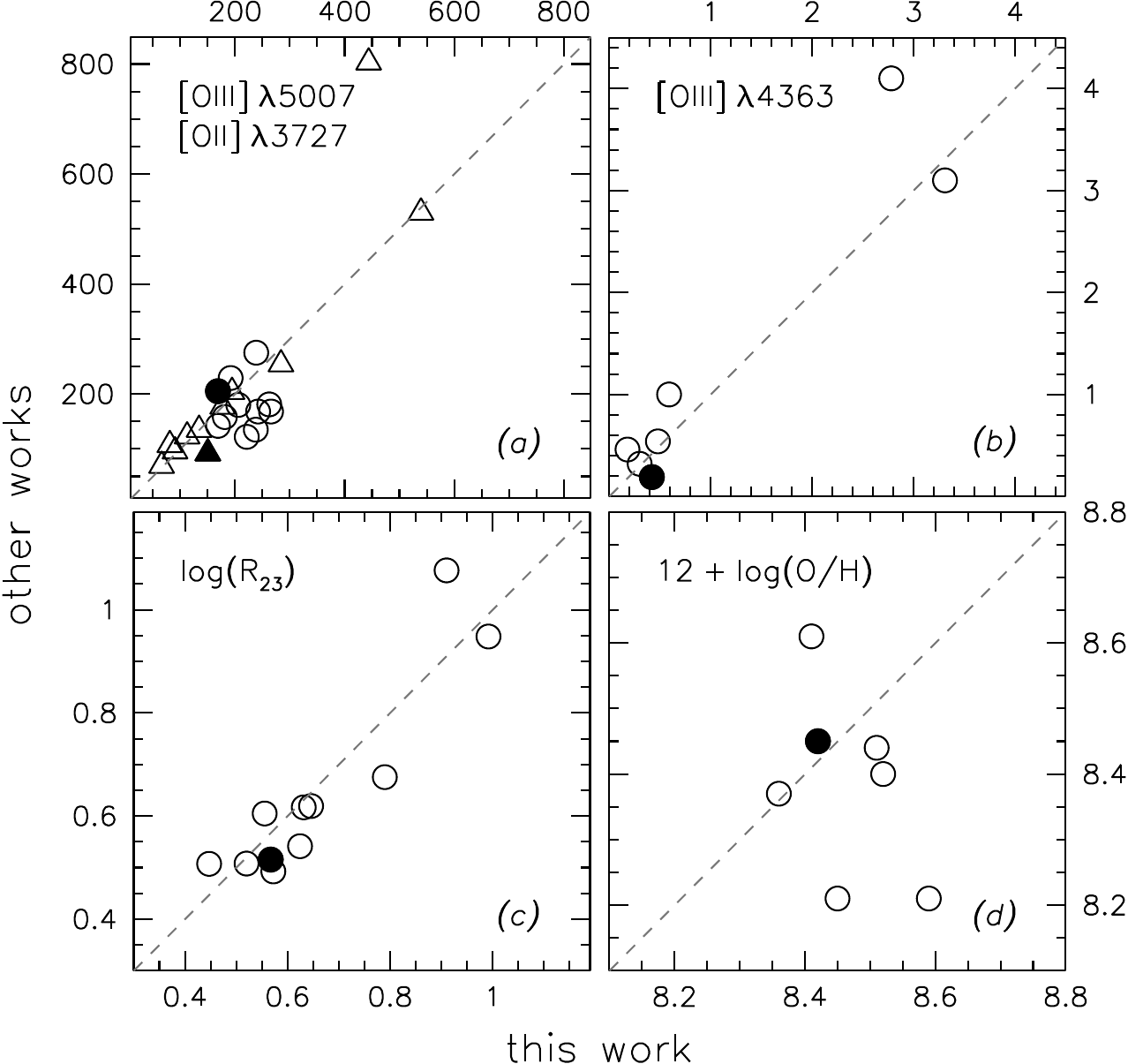}
\caption{Comparison between the data obtained for this work and results from the literature for: {\em (a)} 
 the \oii\lin3727 {\em (circles)} and \oiii\lin5007 {\em (triangles)} reddening-corrected fluxes (in units of H$\beta=100$); {\em (b)} 
 the \oiii\lin4363 fluxes; {\em (c)} log(\rtwothree); {\em (d)} \oh.
   The dashed line represents the one-to-one correlation line. The open and full symbols refer to the comparison with \rs\ and \citet[NGC~595 only]{Esteban:2009}, respectively.
   \label{fig:comparison}}
\end{figure}

\section{The radial abundance gradient and its scatter}\
The radial abundance gradient in M33 has been investigated from \hii\ region spectroscopy by several authors in the past few years (see \rs, \citealt{Magrini:2010}, \citealt{Bresolin:2010}, and references therein). The large number of available \hii\ regions with published line flux data  is utilized here to infer the scatter in abundance at a given galactocentric distance. The new Gemini sample was designed to increase the number of observations near the galactic center, where the detection of the auroral lines is most critical for the radial abundance gradient, although problematic even with 8m-class telescopes.

Line fluxes for 83 \hii\ regions with \oiii\lin4363 detections (in addition to \oii\lin3727 and \oiii\lin5007) have been collected from the following papers: \citet{Kwitter:1981}, \citet{Vilchez:1988}, \citet{Crockett:2006}, \rs, \citet{Magrini:2010} and \citet[Subaru observations]{Bresolin:2010}.
The oxygen abundances were recomputed from the published reddening-corrected emission line fluxes with the same procedure and atomic data used for the Gemini \hii\ region sample. The resulting radial abundance gradient is shown in Fig.~\ref{fig:r23-1}, where different symbols are used for different data sources. The large scatter of the data points in the diagram, already noticed by \rs\ for their sample, is evident: the rms scatter around the least square fit (dashed line) for the full sample is 0.17 dex, apparently larger than what can be accounted for by the measuring uncertainties.
However, looking at the distribution of the data points in the central 2.2~kpc of the galaxy, the new Gemini data presented here, together with the Subaru data from \citet{Bresolin:2010}, display a much lower scatter (0.06 dex, 11 objects), compared to the \rs\ data (0.21 dex, 19 objects) pertaining to the same inner portion of the disk. A Kolgomorov-Smirnov test indicates that the probability of obtaining the low scatter measured for the Gemini+Subaru sample
from the distribution of the \rs\ data points is \lt 3\%.
This suggests that the dispersion  in Fig.~\ref{fig:r23-1} might be related to the quality of the spectroscopic data drawn from the literature, for example to the signal-to-noise (S/N) ratio of the \oiii\lin4363 line detections, rather than intrinsic azimuthal variations on scales of the order of a few hundred pc.  It can also be seen that the \rs\ abundances for the two innermost \hii\ regions (B\,43b and B\,29) are significantly lower  ($\sim0.3$ dex) than those measured from the Gemini sample. 

To test the effect of the reported  quality of the \lin4363 line detection on the abundance scatter, \hii\ regions for which this line was observed with a S/N\,$<$\,6 were removed, together with 11 \rs\ objects in common with the Gemini and Subaru observations done by the author. The result is shown in Fig.~\ref{fig:r23-2}. The overall scatter is only slightly reduced to 0.13, still a factor of two larger than what is measured in the inner two kpc from the author's data alone. However, the apparent abundance peak shown by the \rs\ data at 1.5~kpc from the galaxy center, as well as the lack  of high O/H values in the central kpc present in the \rs\ sample, are now gone.
The radial oxygen abundance gradient measured from this subsample (short-dashed line in Fig.~\ref{fig:r23-2}) is:

\begin{equation}
\rm \oh = 8.48~ (\pm 0.04)~ - ~ 0.042~ (\pm 0.010)~R_{kpc}
\end{equation}

\noindent
where $\rm R_{kpc}$ is the galactocentric distance in kpc.
A $2\,\sigma$ iterative clipping was also applied to the data, providing a comparable result, but with smaller errors (long-dashed line):

\begin{equation}
\rm \oh = 8.50~ (\pm 0.02)~ - ~ 0.045~ (\pm 0.006)~R_{kpc}.
\end{equation}

\noindent
Because of the removal of several low-metallicity \hii\ regions in the central two kpc belonging to the \rs\ sample the slope of the gradient given above is larger than the one determined by \rs\ ($-0.027\pm0.012$~dex kpc\expone). The intercept value is also larger, 8.50 \vs\ 8.36.
The measurement of the new Gemini \te-based abundances in the inner two kpc of M33 thus shows that {\em (i)} it is possible that the real abundance scatter in this galaxy is not as large as claimed by \rs;  {\em (ii)} the radial abundance gradient in M33 behaves `normally', as seen in other galaxies, with the metallicity increasing smoothly towards the center, without indications of a central dip, which has been ascribed by \citet{Magrini:2010} to a selection bias. The interpretation given here 
is that it might have to do instead with errors in the measurements. \citet{Magrini:2010} have also proposed that 
using only giant \hii\ regions to determine the abundance gradient yields a steeper gradient than using the full sample, regardless of luminosity. This is not confirmed by the abundance  analysis carried out in the current paper. From the same nine giant \hii\ regions they used, the abundance gradient is:

\begin{equation}
\rm \oh = 8.50~ (\pm 0.10)~ - ~ 0.040~ (\pm 0.024)~R_{kpc},\label{eq:grad}
\end{equation}

\noindent
which is essentially the same as found above. 

The latter result, based on only nine regions, is useful to offer an explanation as to how various authors have come up with different values for the slope 
of the gradient in M33. The 1\,$\sigma$ range for the slope in Eq.~\ref{eq:grad}  virtually encompasses all of the \hii\ region-based values published in the past twenty years, from $-0.012\pm0.011$~dex kpc\expone\ (\citealt{Crockett:2006}) to $-0.05\pm0.01$~dex kpc\expone\ (\citealt{Vilchez:1988}, after scaling to the M33 distance adopted here and removing the central \hii\ regions), several of which have been determined from a similarly small number of $\sim$10 \hii\ regions. (The central abundances from \citealt{Vilchez:1988} are model-based, and lead to a steep gradient in the inner portion of the disk; however, the direct measurements based on the new auroral line detections do not support such high abundance values). A significantly larger number of data points, well distributed in radius, is
required to reduce the uncertainty in the slope (\citealt{Dutil:2001,Bresolin:2009a}).

\medskip
As a final remark,  the spatial distribution of the \hii\ regions in the central parts of M33  is very similar between the Gemini and the \rs\ samples, with the majority  of the objects located along the spiral arms. 
Although it is implicit in the distribution of the Gemini and Subaru data points in Fig.~\ref{fig:r23-2}, it is worth emphasizing that  pairs of \hii\ regions with nearly coincident galactocentric distances, but situated almost diametrically opposite each other relative to the galaxy center have identical oxygen abundances, within the errors. This is the case for B\,90 \vs\ B\,50, B\,72 \vs\ B\,33b and B\,302 \vs\ C\,1Ab (the abundances for the three \hii\ regions studied with Subaru by \citealt{Bresolin:2010} are summarized in Table~\ref{table:Subaru} for convenience). This implies that the ISM of M33 is well mixed azimuthally over scales of at least 4~kpc. 

\input{tab3}

\begin{figure}
\medskip
\epsscale{1.15}
\plotone{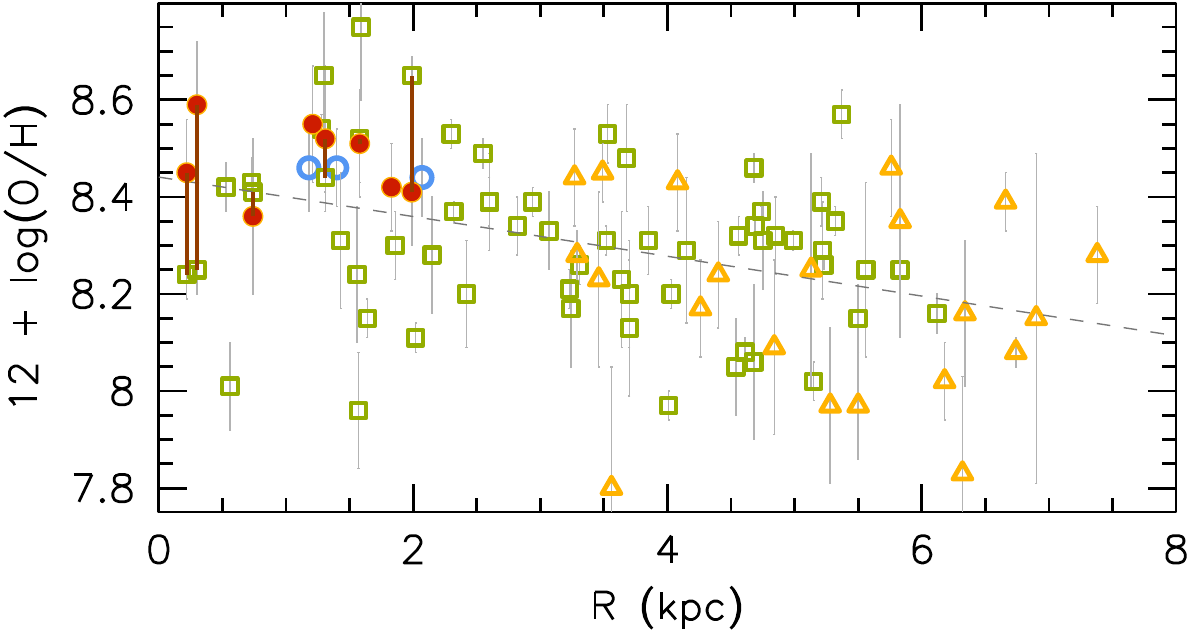}
\caption{Radial oxygen abundance gradient in M33, using \oiii\lin4363-based abundances from this paper (full circles), \citet[open circles]{Bresolin:2010}, \citet[squares]{Rosolowsky:2008} and other sources in the literature (triangles). The least square fit to the full dataset is shown by the dashed line. Vertical segments connect objects in common between the Gemini and \rs\ samples.
\label{fig:r23-1}}
\end{figure}

\begin{figure}
\medskip
\epsscale{1.15}
\plotone{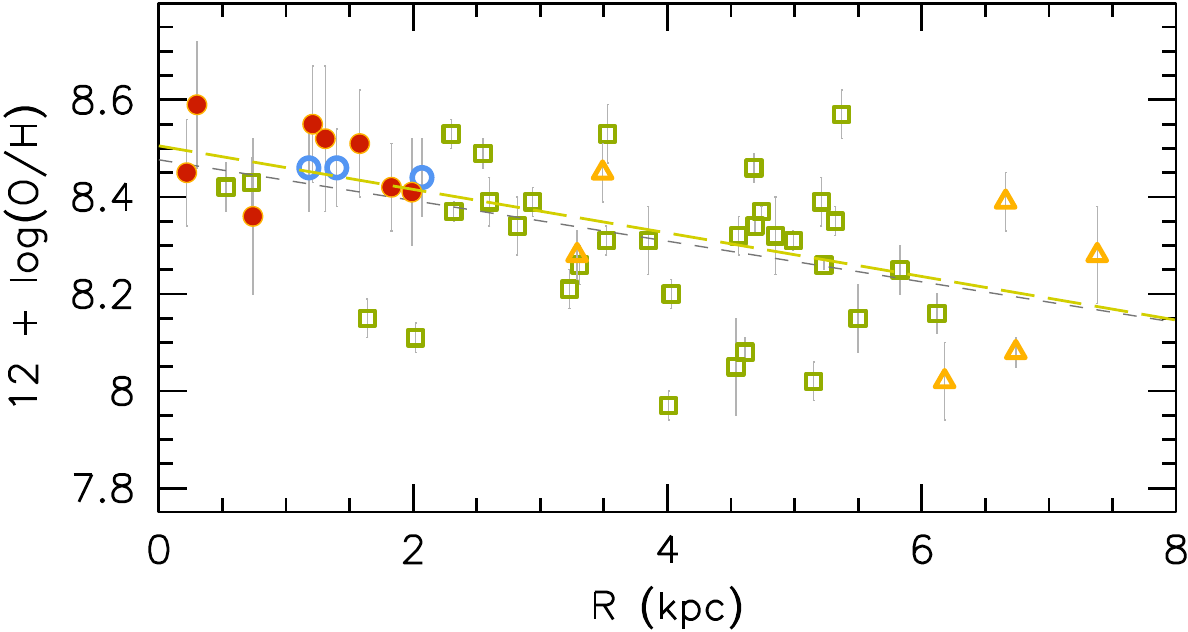}
\caption{Same as Fig.~\ref{fig:r23-1}, with sample reduced to the highest signal-to-noise \oiii\lin4363 detections, as explained in the text.
The lines represent fits to all the data points shown (short dashes) and to a smaller sample obtained by means of a $2\,\sigma$ iterative clipping procedure (long dashes).
\label{fig:r23-2}}
\end{figure}

\subsection{The abundance scatter determined from \rtwothree} 
Despite the fact that the subsample of \hii\ regions selected from the literature with a reported detection of \oiii\lin4363 better than 6\,$\sigma$ displays only a slight improvement in the rms scatter around the exponential gradient compared to the full sample, the much smaller dispersion
from the Gemini+Subaru observations by the author suggests that the real abundance fluctuations in M33 are actually smaller and comparable to what is observed in other well-studied spiral galaxies.
To test this assertion, and to show that it not simply a consequence of the relatively small number  of objects considered (11) in the inner disk alone,
the oxygen abundances for the \hii\ region sample shown in Fig.~\ref{fig:r23-1} were computed with the
\rtwothree\ abundance indicator, which uses only strong emission lines, so that the resulting abundances are largely independent of the random errors affecting the much weaker  \oiii\lin4363 auroral lines, and which propagate through the chemical abundance measurement.
Despite the large systematic uncertainties affecting \rtwothree\ and other strong-line methods, as shown by the comparison of \te-based abundances with different calibrations of strong-line indicators (e.g.~\citealt{Bresolin:2009a}), the abundance scatter should be comparable to what high-quality auroral line detections indicate (somewhat larger dispersions can be expected from the dependence of \rtwothree\ on the ionization conditions of the different \hii\ regions, see \citealt{Kennicutt:1996}). 
In fact, as mentioned earlier, \citet{Bresolin:2009a} measured the abundance gradient of NGC~300 from \oiii\lin4363 line detections in 28 \hii\ regions, finding an rms scatter of 0.05 dex. Comparable values, between 0.06 and 0.08 dex, are measured  from the same data set when the abundances are obtained from a variety of strong-line abundance indicators.

\medskip
The \rtwothree\ calibration of \citet{McGaugh:1991}, as given in \citet{Kuzio-de-Naray:2004}, was adopted, since it has the advantage over other available calibrations of accounting 
for the nebular gas excitation, as given by the $y$\eq log(O3/O2) parameter, with O3\eq \oiii\llin4959,\,5007/\hbeta\ and 
O2\eq \oii\lin3727/\hbeta. The upper branch calibration was used, considering the published \nii/\halpha\ values, and the position
of the targets in a \rtwothree\ vs O/H plot (presented in Sec.~\ref{section:systematic}). 

The O/H abundance ratio thus calculated is shown as a function of radial distance in Fig.~\ref{fig:r23grad1} for the same \hii\ regions presented in  Fig.~\ref{fig:r23-1}.
A subsample, shown with full symbols,  was created by selecting objects with {\em (a)} S/N (\oiii\lin4363)\gt 5; {\em (b)} 
\neiii\lin3868/\hbeta \lt 0.45; {\em (c)} \oiii\lin5007/\hbeta \lt 5; {\em (d)} log\rtwothree \lt 0.90.
These criteria will be justified and discussed at length in Sec.~\ref{section:systematic}, and serve to isolate \hii\ regions that are not affected by
extremely hard ionization fields, or objects for which the \rtwothree-based abundances could be very uncertain.
\hii\ regions that do not satisfy these conditions are shown either with open symbols (low S/N in the \oiii\lin4363 line), double open symbols (strong \neiii\ and \oiii\ lines) and crossed symbols (large \rtwothree). 

The dashed line in Fig.~\ref{fig:r23grad1} represents a least square fit to the data belonging to the subsample thus obtained, limiting the fit to galactocentric distances smaller than 5 kpc (to avoid possible biases when reaching down to metallicities near the turnover region of \rtwothree, as explained in Sec.~\ref{section:systematic}). The result is

$$\rm \oh = 8.82~ (\pm 0.02)~ - ~ 0.033~ (\pm 0.005)~R_{kpc}.$$ 

\noindent
The slope is slightly shallower  than the value obtained from the auroral lines, although the two are still consistent within the 1\,$\sigma$ errors.
It is worth recalling here the dependence of the slope of galactic radial abundance gradients  on the method chosen to measure the oxygen 
abundances for their \hii\ regions (e.g.~\citealt{Bresolin:2009a}). As an example, applying the widely-used \rtwothree\ calibrations by \citet{Tremonti:2004} and \citet{Zaritsky:1994} yields slopes of $-0.044 \pm 0.006$ dex\,kpc\expone\ and $-0.051 \pm 0.007$ dex\,kpc\expone, respectively.
Moreover, as mentioned earlier, there are important systematic discrepancies in the metallicities derived from strong-line methods calibrated from theoretical model grids (as in the case of the \rtwothree\ calibration by 
\citealt{McGaugh:1991} adopted here)  and the auroral-line method. This explains the large offset between the intercepts of the regression lines shown in Fig.~\ref{fig:r23-1} and Fig.~\ref{fig:r23grad1}.
However, this has virtually no effect on the derivation of the abundance scatter.

The most important result to note is that the rms scatter of the data shown with full symbols in Fig.~\ref{fig:r23grad1}
is only 0.05 dex, which is virtually the same value found from the \oiii\lin4363-based abundances for the innermost \hii\ regions observed with Gemini and Subaru. Changing the \rtwothree\ calibration has virtually no effect, as the rms scatter remains between 0.05 and 0.07 dex throughout the disk of the galaxy.
Following \rs, the least-square method by \citet{Akritas:1996}, which computes the {\em intrinsic} scatter  of the data, was applied to the \rtwothree-based abundances, confirming that the observed scatter is due only to measurement errors. The data therefore indicate the absence of real azimuthal variations in the oxygen abundances.

It should be pointed out that the \rs\ \hii\ regions located in the central two kpc, which showed a very large scatter in the \te-based abundance plot of Fig.~\ref{fig:r23-1}, are now in excellent agreement with the new Gemini data, and have a comparably small abundance scatter. It should also be added that the selection based on the S/N of the \oiii\lin4363 line has little effect on the measured rms scatter, but it helps to retain the targets with the best determined values of the \oii\lin3727 and \oiii\llin4959,\,5007 lines, used to define the \rtwothree\ indicator.

\medskip
The main conclusion that can be drawn from the analysis of the \rtwothree-based abundances, supported by the discussion of the \te-based
results in the inner two kpc,  is that the abundance scatter in M33 is small (0.05-0.06 dex), comparable to the measuring errors, and similar in magnitude to what is observed in other well-studied spiral galaxies, such as NGC~300. 
We can therefore speculate that the significant intrinsic scatter seen in the \te-based abundances of M33 from the \rs\ sample is due to measurement errors in the weak \oiii\lin4363 auroral line. From Fig.~\ref{fig:r23-2} and \ref{fig:r23grad1} it is also possible to claim that the oxygen abundance reaches its peak at the center of the galaxy, as is typically the case for spiral galaxies, and therefore  that there is no evidence for the off-centered abundance distribution suggested by \citet{Magrini:2010}.

\begin{figure}
\medskip
\epsscale{1.15}
\plotone{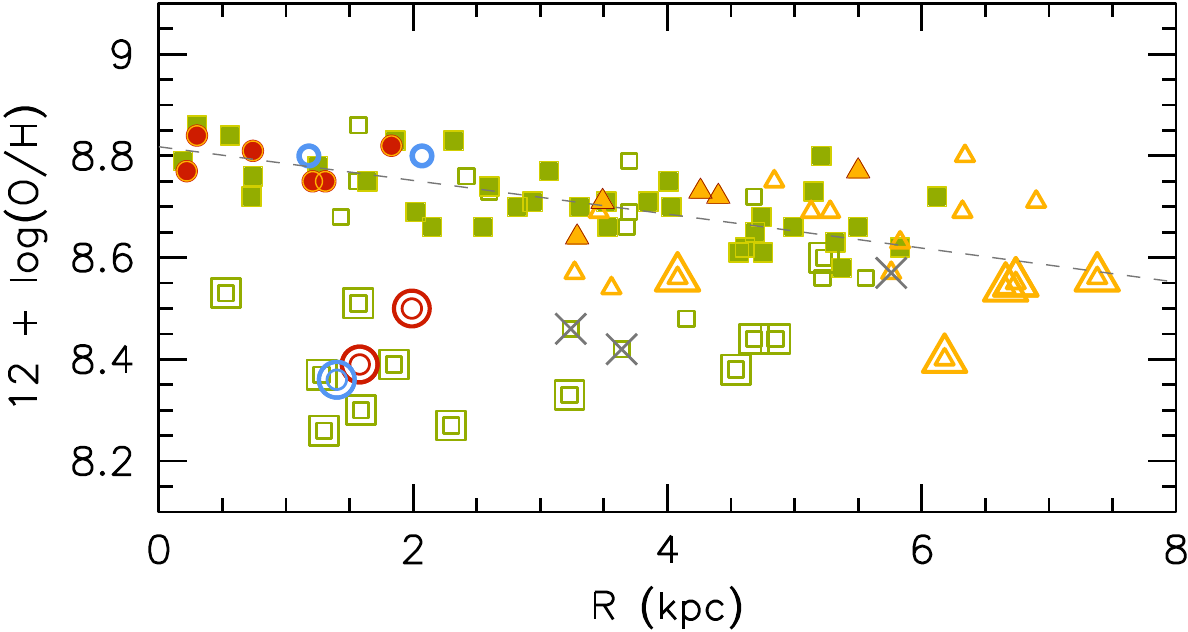}
\caption{Radial abundance gradient in M33 measured from the \rtwothree\ indicator. \hii\ regions from the new Gemini observations (red circles), the Subaru data of \citet[bue circles]{Bresolin:2010}, the sample of \rs\ (squares) and other data from the literature (triangles) are included.  The dashed line represents the linear fit to the subsample defined in the text and shown by full symbols.
Double-lined symbols are used for \hii\ regions with very high \neiii\ and \oiii\ fluxes. Crossed symbols are used for additional objects with log(\rtwothree) \gt\ 0.90. The remaining open symbols represent \hii\ regions with low S/N in the \oiii\lin4363 line.
\label{fig:r23grad1}}
\end{figure}

\section{Systematic effects of a hard ionizing radiation on the \rtwothree\ abundances}\label{section:systematic}
In Fig.~\ref{fig:r23grad1} a number of targets   were found to lie systematically below the sequence of \hii\ regions defining the \rtwothree-based abundance gradient in M33. In particular, in the inner part of the disk  the  oxygen abundances of several \hii\ regions  are $\sim$0.3 to 0.5 dex below the regression representing the abundance gradient, and appear  to belong to a secondary sequence of data points that merges into the main one at a galactocentric distance of about 7~kpc.

In order to interpret the origin for this discrepancy, and justify the exclusion of these targets from the \rtwothree-based abundance gradient analysis, the \te-based oxygen abundances of the same \hii\ regions included in Fig.~\ref{fig:r23grad1} are shown as a function of \rtwothree\ in Fig.~\ref{fig:ohr23}.
In this diagram, a number of  extragalactic \hii\ regions with high-quality  metallicity determinations, based on the \oiii\lin4363 auroral line, are also included (small dots). They were drawn  from the following sources:\\

\noindent
\begin{tabular}{ll}
NGC~300:						&		\citet{Bresolin:2009a}\\[0.5mm]
M101: 							&		\citet{Kennicutt:2003}\\[0.5mm]
NGC~2403:						&		\citet{Garnett:1997}\\
								&		\citet{Esteban:2009}\\[0.5mm]
IC~1613: 							&		\citet{Bresolin:2006a}\\[0.5mm]
\hii\ galaxies: 						&		\citet{Izotov:1994}\\
								&		\citet{Guseva:2009}\\[0.5mm]
M31, NGC~2363, NGC~1741			&		\citet{Esteban:2009}\\					
NGC~4395, NGC~4861				&	\\
\end{tabular}

\medskip
\noindent
These data  illustrate the well-known bi-valued nature of \rtwothree, consisting of two \hii\ region sequences, intersecting around \oh\eq8.0. The curves in Fig.~\ref{fig:ohr23} are polynomial fits shown to guide the eye in recognizing the two branches. The \hii\ region data for M33 are plotted with the same symbols used in Fig.~\ref{fig:r23grad1}. It can be immediately seen that the majority of the most discrepant points in Fig.~\ref{fig:r23grad1}, identified by the large doubled-lined symbols, lie at large \rtwothree\ values and above the intersection between the upper and lower branches in Fig.~\ref{fig:ohr23}.
Very large \rtwothree\ values (log \rtwothree \gt 1) are relatively rare among extragalactic \hii\ regions, and are encountered in  high-excitation nebulae with large ionization parameters.  A fraction of such objects  show spectral features  associated with the presence of massive stars
emitting particularly hard ionizing radiation, such as early WN (WNE)  stars (\citealt{Crowther:2007}). It is thus not surprising that the \hii\ regions where \rs\ have identified nebular \heii\lin4686 line emission (I.P.\eq54.4 eV) are located, together with additional discrepant points, in the upper right corner of the \rtwothree\ \vs\ O/H plot. Invariably in this zone of the diagram, even when the \heii\ emission remains undetected or the ionizing sources cannot be unequivocally identified, the high-excitation lines \neiii\lin3868 (I.P.\eq 41 eV) and \oiii\lin5007 (I.P.\eq 35 eV) are particularly strong. 

\begin{figure}
\medskip
\epsscale{1.15}
\plotone{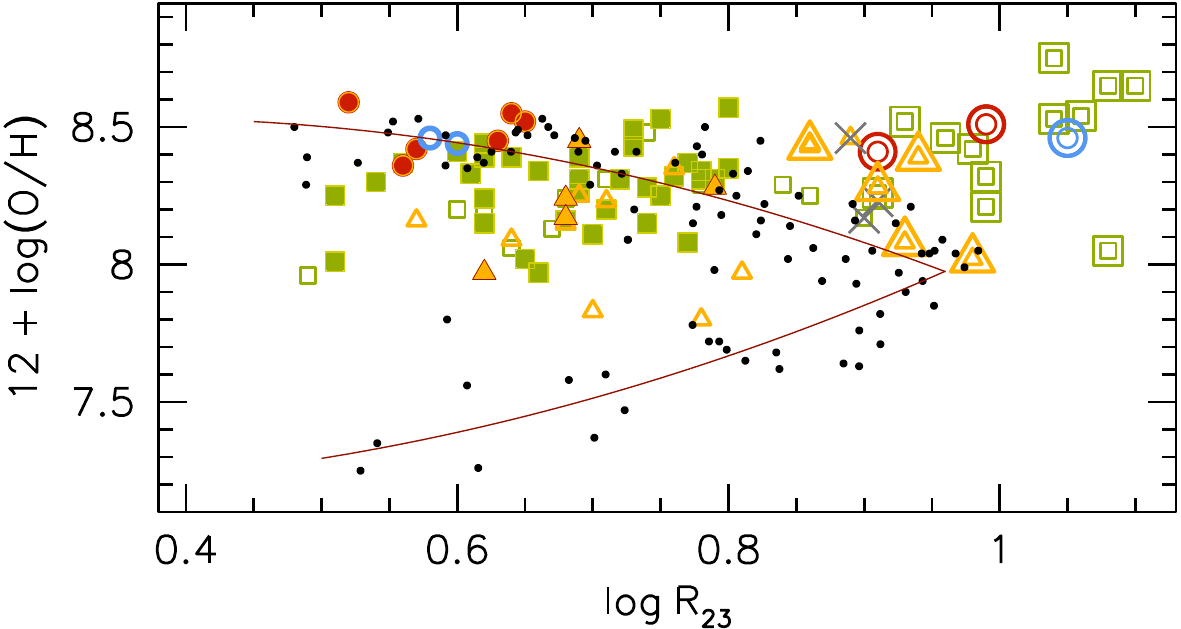}
\caption{Oxygen abundances derived from the \oiii\lin4363 auroral line as a function of log(\rtwothree). A sample of \hii\ regions with high-quality measurements in different spiral and irregular galaxies (small dots) have been used to schematically define the upper and lower branches of \rtwothree\ (shown by the two intersecting curves). \hii\ regions from the new Gemini observations (red circles), the Subaru data of \citet[bue circles]{Bresolin:2010}, the sample of \rs\ (squares) and other data from the literature (triangles) are included. The symbols used are the same as in Fig.~\ref{fig:r23grad1}.\label{fig:ohr23}}
\end{figure}

\subsection{\hiiit\ regions with large \neiii\ and \oiii\ line fluxes}
The large double-lined symbols in Fig.~\ref{fig:ohr23} were used to identify
\hii\ regions where \nethree \eq \neiii  \lin3868 /\hbeta \gt 0.45 or \othree\eq\oiii\lin5007/\hbeta \gt 5. 
The use of both line ratios is somewhat redundant, since the two are strongly correlated (the relative values of the criteria adopted for Ne$_{3868}$ and O$_{5007}$ are consistent with the trend found between the two line ratios from the enlarged \hii\ region sample analyzed here).
The choice of the adopted selection criteria  is purely empirical and somewhat arbitrary, and was guided by the knowledge of the line fluxes observed in nearby nebulae ionized by  WNE stars. For instance, in RCW\,5 and RCW\,48 (Milky Way), and N57\,C and  N138\,D,\,B (LMC), \citet{Kennicutt:2000} measured \nethree\gt0.56 and \othree\gt 5.8, while, for example, for the giant \hii\ region 30~Dor in the LMC, ionized by a cluster of O stars, \nethree\eq0.40 and \othree\eq4.3.

Typical {\em upper branch} \hii\ regions in other galaxies are observed with \nethree\lt0.40 and \othree\lt 4.0, and highly excited \hii\ regions, such as \#15 (C\,1Ab) and \#19 (B\,23) in Table~\ref{table:sample}, easily stand out among objects with much lower \nethree\ and \othree\ values. However, it should be recalled that high \nethree\ and \othree\ values, also exceeding the limits adopted above, can be encountered in {\em lower branch} \hii\ regions (low metallicities), or near the turnaround between the two \rtwothree\ branches. This explains why a few
\hii\ regions at galactocentric distances \gt 6~kpc were rejected in Fig.~\ref{fig:r23grad1} by the criteria adopted above, even though their abundances
  can agree with the  abundance gradient defined by the remaining regions. In fact, 
the \hii\ regions at large radial distances are near or at the turnaround, as Fig.~\ref{fig:ohr23} shows. For this reason the linear fit to the \rtwothree-based abundances has been limited to galactocentric distances smaller than 5~kpc. 

It should be noted that not all of the discrepant objects in M33 contain known Wolf-Rayet stars, and also that 
the presence of WNE stars in an extragalactic \hii\ region does not necessarily imply 
peculiarly large \nethree\ and \othree\ values. This occurs only if the nebular ionization is dominated by the radiation emitted by the WN star(s), and the dilution resulting from the presence of O stars is small. This agrees with the observation that the total H ionizing flux output of the discrepant M33 \hii\ regions analyzed here,  estimated from the \halpha\ luminosity, is consistent with the ionization being provided by only 1-2 WNE stars. What could be a little surprising is  finding such a large fraction ($\sim$ 20\% of the nebulae with \oiii\lin4363 detections) of highly excited \hii\ regions in M33. To the author's knowledge, for no other galaxy has an
\hii\ region population been insofar found with such a large percentage of nebulae having log(\rtwothree)\gt 1. On the other hand, the typical luminosity of \hii\ regions selected in extragalactic abundance studies is generally significantly higher than that of the majority of those observed in M33. Consequently, their integrated spectra would be less affected by the presence of a few sources of hard radiation amid a large number of ionizing O stars.
In fact, none of the M33 \hii\ regions  with a total luminosity L(\halpha)\gt $10^{38}$ erg\,s\expone\ (adopting fluxes published by \citealt{Hodge:2002} and \citealt{Relano:2009}) for which a spectrum is available displays notable levels of excitation. The only (marginal) exception in the upper branch regime is IC~131 (\nethree\eq0.47, \othree\eq3.9, \citealt{Vilchez:1988}). The reason in this case is possibly connected to the fact that, besides containing three WN stars, this giant \hii\ region has also a peculiarly hard and extended X-ray emission, whose source is at present unexplained (\citealt{Tullmann:2009}).

\begin{figure}
\medskip
\epsscale{1.15}
\plotone{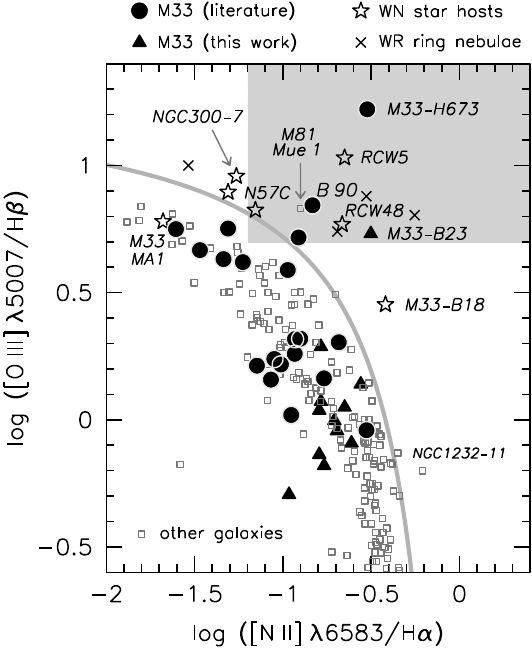}
\caption{The \oiii/\hbeta\ \vs\ \nii/\halpha\ diagnostic diagram for \hii\ regions in M33 from this work (triangles) and from the literature (full circles). \hii\ regions in the Milky Way and the Large Magellanic Cloud (\citealt{Kennicutt:2000}), NGC~300 (\citealt{Bresolin:2009a}) and M33 (this work and \citealt{Kehrig:2010}) hosting WN stars are included (star symbols; most are identified). 
WR ring nebulae from \citet{Esteban:1993} are shown with crosses.
Additional nebulae contained in spiral and irregular galaxies are taken from \citet{Bresolin:1999, Bresolin:2005} and other sources (open squares) to help discern the \hii\ region excitation sequence.
The curve is the boundary between star-forming and AGN-dominated galaxies defined by \citet{Kauffmann:2003}. The shadowed area includes high-excitation \hii\ regions in the upper \rtwothree\ branch having 
\oiii/\hbeta\ \gt\ 5.
\label{fig:bpt}}
\end{figure}

\subsection{Excitation diagnostic diagrams}
To clarify the effect of the presence of WNE stars or other sources of hard ionizing radiation on the emission line spectra of \hii\ regions,
Fig.~\ref{fig:bpt} shows a \nii\lin6583/\halpha\ \vs\ \oiii\lin5007/\hbeta\ diagnostic diagram (\citealt{Baldwin:1981}), in which M33 \hii\ region data from the literature (including the recent work on \heii-emitting \hii\ regions by \citealt{Kehrig:2010}) are shown together with a small number of Galactic and extragalactic  \hii\ regions (\citealt{Kennicutt:2000, Bresolin:2009a}) containing identified WNE stars (star symbols). 
The WR ring nebulae G2.4\,+\,1.4 (WO1), NGC~2359 (WN4), NGC~3199 and S308 (WN5) from \citet{Esteban:1993} are also shown (crosses).
Additional \hii\ region data for spiral and irregular galaxies from the literature (\citealt{Bresolin:1999,Bresolin:2005}, and those shown as small dots in Fig.~\ref{fig:ohr23}) are included for comparison. 
Nebulae lying above the curve (taken from \citealt{Kauffmann:2003}) are considered to be ionized by a harder ionizing field than provided by O stars. 
In this plot the selection of  objects with \othree\gt 5 for the discrepant points in Fig.~\ref{fig:ohr23} corresponds to 
the shaded area, assuming that log(\nii\lin6583/\halpha) \gt $-1.2$ on the upper branch (\citealt{Kewley:2008}).

It can be seen that the presence of WNE stars can shift \hii\ regions into the grey area and above the boundary line. Still, there can be low-metallicity (small \nii/\halpha\ ratio) nebulae containing WNE stars (e.g.~MA\,1 in M33) whose position in the diagram agrees with the excitation sequence defined by the control sample. Moreover, some of the high-excitation \hii\ regions have no clear detection of 
associated hot stars  (e.g.~H\,673  in M33, \citealt{Kehrig:2010}).
For 13 objects in the Gemini sample the spectral coverage extended to sufficiently long wavelengths that the \nii\lin5755 could be measured. From the knowledge of the electron temperature or assuming \te\eq$10^4~K$ the strength of the \nii\lin6583 was estimated
using the {\tt\small temden} program in {\sc iraf}.
 These objects were then added to Fig.~\ref{fig:bpt} as triangles. One of these, region B\,18, contains a WN star (\citealt{Massey:1998}) clearly detected in the Gemini spectrum from its broad \heii\lin4686 emission line. This \hii\ region lies to the right of the 
boundary drawn in the diagram.
For another, region B\,23, situated into the shaded area, no possible match
in the Wolf-Rayet catalog by \citet{Massey:1998} could be found (the same is true for B\,90, observed with the Subaru telescope),
although its excitation, as measured from the \oiii/\hbeta\ line ratio, is higher than for B\,18.

The position of some of the deviating \hii\ regions can be linked to the presence of shocks, which can enhance the strength of the \nii\ emission line. This is almost certainly the case for region 11 in NGC~1232 (\citealt{Bresolin:2005}, identified near the bottom of
Fig.~\ref{fig:bpt}), whose spectrum shows strong \oi\lin6300 and \sii\llin6717,\,6731 emission (thus being a supernova remnant candidate). For other \hii\ regions the situation is less clear. For example, H\,673 in M33 has moderately strong \nii\ and \sii\ fluxes, but no \oi\ emission has been reported by \citet{Kehrig:2010}. Nitrogen self-enrichment is also an option to explain the large \nii\ flux, since these authors report log(N/O)\eq$-0.74\pm0.28$, versus a typical log(N/O)\eq$-1.2$  for \hii\ regions in M33 (\citealt{Bresolin:2010}).

Fig.~\ref{fig:bpt} shows that also M\"unch~1, an outlying object  in M81, studied in some detail by \citet{Garnett:1987}, is among the restricted number of high-excitation extragalactic \hii\ regions contained within the shaded area (\nethree\eq0.46, \othree\eq6.9).
 Given its location in the BPT diagram, it is possible that future high-sensitivity spectra of this region will reveal nebular \heii\ and/or WN-related stellar emission features.

\subsection{\hiiit\ regions with large \rtwothree\ values}
In Fig.~\ref{fig:r23grad1} it was decided to remove from the final sample also the \hii\ regions for which log(\rtwothree)\gt 0.9. The crossed symbols in the plot represent the three
\hii\ regions that were removed based only on this condition (rather than poor S/N, or large \neiii\ and \oiii\ line fluxes).
The condition on \rtwothree\ is added because 
the turnover in the oxygen abundance \vs\ \rtwothree\ relation occurs around log(\rtwothree)\,$\simeq$\,1. It is well known that here
the \rtwothree\ indicator is a poor abundance diagnostic, because a wide range of abundances corresponds to a narrow range
of indicator values (\citealt{McGaugh:1991}). Studies of radial abundance gradients in spiral galaxies should therefore exclude \hii\ regions in the turnover region in order to avoid abundance biases or large measuring uncertainties.
It should also be noted that the selection based on \rtwothree\ effectively removes virtually all of the targets with high \neiii\ and \oiii\ fluxes. The previous discussion, however, is helpful to understand what kind of objects systematically deviate from
the \rtwothree\ \vs\ O/H trend of Fig.~\ref{fig:ohr23}.

\medskip
In conclusion, \hii\ regions ionized by uncommonly  hard ionizing radiation (such as that generated by the presence of early WN stars, but other  poorly constrained processes are  also at work) are shifted to higher \rtwothree\ values at constant O/H abundance ratios compared to `normal' \hii\ regions in a diagram like the one shown in 
Fig.~\ref{fig:ohr23}. As a consequence, \rtwothree\  would systematically underestimate the abundances for these objects, as observed in Fig.~\ref{fig:r23grad1}, because the upper branch bends to smaller O/H values with increasing \rtwothree. This also explains the secondary sequence seen
at the bottom of this figure, progressively diverging from the gradient regression line with decreasing radial distance (equivalent to increasing \rtwothree).
The empirical emission-line criteria outlined above allow to discriminate quite effectively against such discrepant \hii\ regions.

\section{Summary}
New GMOS/Gemini spectra of 25 \hii\ regions located in the central two kpc of M33 have been analyzed. The oxygen abundances have been derived from the \oiii\lin4363 auroral line for eight of the targets. The scatter in the radial abundance gradient has been investigated by combining 
 existing samples of \hii\ regions in this galaxy  with the new data. The main conclusions of this work are summarized as following:

\begin{trivlist}

\item {\em (1)} the scatter in the oxygen abundance in the central two kpc, as derived from the auroral line measurements for the new Gemini sample, is approximately 0.06 dex, much lower than the value of $\sim0.21$ dex obtained in the same inner portion of the disk from the \hii\ regions observed by \citet{Rosolowsky:2008}.

\item {\em (2)} the oxygen abundances estimated from the \rtwothree\ metallicity indicator for
a large sample of \hii\ regions assembled from the literature in combination with the new observations yield a comparably small scatter (0.05-0.07 dex) {\em over the whole optical disk of M33}. This dispersion can be explained simply by the measuring uncertainties.

\item {\em (3)} no evidence is found for significant {\em intrinsic} azimuthal variations in the \hii\ region metallicity distribution in M33 on scales from $\sim$100 pc to a few kpc, as well as for a displacement of the abundance peak from the galaxy center. This result rules out large anomalies in the mixing of the gas.

\item {\em (4)} a considerable fraction of the M33  \hii\ regions with auroral line detections  show spectral features revealing sources of hard ionizing radiation (\heii\ emission, large \neiii\ and \oiii\ line fluxes). In some cases these can be identified with early WN stars, but several cases exist where no obvious source has been identified.
For these nebulae the oxygen abundances derived from \rtwothree\  are systematically underestimated by large factors. However, it is straightforward to identify these highly excited \hii\ regions from their large \rtwothree\ values and their position in the excitation diagnostic diagrams. Thus,  abundance gradients derived for other spiral galaxies hosting such objects  would still be correct, once these peculiar nebulae are removed from the abundance analysis based on the strong-line metallicity indicator.

\end{trivlist}

\acknowledgments
FB gratefully acknowledges the support from the National Science Foundation grants AST-0707911 and AST-1008798, and thanks the referee, Cesar Esteban, for comments that helped to improve the manuscript.
The Gemini observations used in  this paper were obtained under program GN-2009B-Q-5.

\medskip
\noindent
{\it Facility:} \facility{Gemini:Gillett (GMOS)}\\

\bibliography{m33}

\clearpage

\end{document}

%% file: tab1.tex
\begin{deluxetable}{ccccl}
\tabletypesize{\scriptsize}
\tablecolumns{5}
\tablewidth{0pt}
\tablecaption{Observed H\,\scriptsize II \small region sample\label{table:sample}}

\tablehead{
\colhead{\phantom{aaa}ID\phantom{aaa}}	     &
\colhead{\phantom{aaaa}R.A.\phantom{aaaa}}	 &
\colhead{\phantom{aaaa}Decl.\phantom{aaaa}}  &
\colhead{\phantom{aaa}R\phantom{aaa}}	 &
\colhead{\phantom{aaa}other ID\phantom{aaa}}	 \\[0.5mm]
\colhead{}       &
\colhead{(J2000.0)}   &
\colhead{(J2000.0)}   &
\colhead{(kpc)}   &
\colhead{} \\[1mm]
\colhead{(1)}	&
\colhead{(2)}	&
\colhead{(3)}	&
\colhead{(4)}	&
\colhead{(5)}	}
\startdata
\\[-2mm]

1\dotfill   &      1~ 33~ 33.53	&	 30~ 41~ 29.8   &    1.83   &	NGC~595\\   
2\dotfill   &      1~ 33~ 39.94	&	 30~ 41~ 07.4   &    1.21   &	B\,50\\   
3\dotfill   &      1~ 33~ 43.69	&	 30~ 40~ 58.1   &    0.87   &	B\,52\\   
4\dotfill   &      1~ 33~ 49.65	&	 30~ 39~ 55.6   &    0.17   &	H\,670\\   
5\dotfill   &      1~ 33~ 48.39	&	 30~ 39~ 35.7   &    0.22   &	B\,43b$^*$\\   
6\dotfill   &      1~ 33~ 48.17	&	 30~ 39~ 16.4   &    0.22   &	B\,37\\   
7\dotfill   &      1~ 33~ 36.31	&	 30~ 38~ 48.4   &    1.22   &	B\,1008a\\   
8\dotfill   &      1~ 33~ 47.78	&	 30~ 38~ 37.6   &    0.30   &	B\,29$^*$\\   
9\dotfill   &      1~ 33~ 49.42	&	 30~ 38~ 00.2   &    0.42   &	B\,20a\\   
10\dotfill 	&      1~ 33~ 50.07	&	 30~ 37~ 30.2   &    0.57   &	B\,16$^*$\\   
11\dotfill 	&      1~ 33~ 34.86	&	 30~ 37~ 05.2   &    1.31   &	B\,33b$^*$\\   
12\dotfill 	&      1~ 33~ 45.48	&	 30~ 36~ 48.5   &    0.74   &	B\,27a$^*$\\   
13\dotfill 	&      1~ 34~ 01.78	&	 30~ 35~ 49.7   &    1.71   &	B\,1$^*$\\   
14\dotfill 	&      1~ 33~ 59.32	&	 30~ 35~ 46.5   &    1.55   &	B\,4a\\   
15\dotfill 	&      1~ 34~ 00.32	&	 30~ 34~ 17.1   &    1.99   &	C\,1Ab$^*$\\   
16\dotfill 	&      1~ 34~ 00.03	&	 30~ 33~ 56.5   &    2.07   &	B\,8Ac\\   
17\dotfill 	&      1~ 33~ 50.23	&	 30~ 33~ 47.5   &    1.62   &	B\,15a\\   
18\dotfill 	&      1~ 33~ 47.66	&	 30~ 33~ 37.6   &    1.60   &	B\,18$^*$\\   
19\dotfill 	&      1~ 33~ 42.91	&	 30~ 33~ 30.1   &    1.58   &	B\,23$^*$\\   
20\dotfill 	&      1~ 33~ 54.13	&	 30~ 33~ 09.6   &    1.96   &	B\,13c$^*$\\   
21\dotfill 	&      1~ 33~ 47.87	&	 30~ 33~ 05.0   &    1.75   &	B\,17d\\   
22\dotfill 	&      1~ 33~ 59.94	&	 30~ 32~ 45.1   &    2.38   &	B\,703\\   
23\dotfill 	&      1~ 33~ 57.43	&	 30~ 32~ 42.3   &    2.25   &	M\,em\,43\\   
24\dotfill 	&      1~ 33~ 46.92	&	 30~ 32~ 35.1   &    1.87   &	near C\,53\\   
25\dotfill 	&      1~ 33~ 44.52	&	 30~ 32~ 01.1   &    1.98   &	B\,1502a   

\enddata
\tablecomments{Units of right ascension are hours, minutes and seconds, and units of declination
are degrees, arcminutes and arcseconds. Col.~(1): \hii\/ region identification. Col.~(2): Right
ascension. Col.~(3): Declination. Col.~(4): Deprojected galactocentric distance in kpc.
Col.~(5): Identification from \citet[\eq B]{Boulesteix:1974}, \citet[\eq H]{Hodge:1999}, \citet[\eq C]{Courtes:1987} and
\citet[\eq M]{Magrini:2000}. Objects with a $*$ symbol in column (5) are in common with \citet{Rosolowsky:2008}.}
\end{deluxetable}


%% file: tab2.tex
\begin{deluxetable*}{ccccccccccc}
\tabletypesize{\scriptsize}
\tablecolumns{11}
\tablewidth{0pt}
\tablecaption{Reddening-corrected line fluxes and oxygen abundances\label{table:fluxes}}

\tablehead{
\colhead{\phantom{aaaaa}ID\phantom{aaaaa}}	     &
\colhead{\oii}	 &
\colhead{\neiii}	 &
\colhead{\oiii}	 &
\colhead{\hi}	 &
\colhead{\oiii}	 &
\colhead{\oiii}	  &
\colhead{F(H$\beta$)} &
\colhead{$c$(\hbeta)} &
\colhead{\te}    &
\colhead{\oh}	\\[0.5mm]
\colhead{}       &
\colhead{3727}   &
\colhead{3868}   &
\colhead{4363}   &
\colhead{4101}   &
\colhead{4959}   &
\colhead{5007}   &
\colhead{(erg\,s$^{-1}$\,cm$^{-2}$)}     &
\colhead{(mag)}  &
\colhead{(K)}  &
\colhead{} 	   \\[1mm] 
\colhead{(1)}	&
\colhead{(2)}	&
\colhead{(3)}	&
\colhead{(4)}	&
\colhead{(5)}	&
\colhead{(6)}	&
\colhead{(7)}	&
\colhead{(8)}	&
\colhead{(9)}	&
\colhead{(10)}	&
\colhead{(11)}	}
\startdata
\\[-2mm]

 1\dotfill &   169 $\pm$   17 &     4.7 $\pm$  0.5 &    0.42 $\pm$ 0.04 &      23 $\pm$    2 &     50 $\pm$    3 &    150 $\pm$    9 &    1.7 $\times 10^{-14}$ &  0.46 &   7960 $\pm$    300 &    8.42 $\pm$    0.09  \\
 2\dotfill &   213 $\pm$   20 &     7.2 $\pm$  0.7 &    0.42 $\pm$ 0.07 &      25 $\pm$    2 &     56 $\pm$    3 &    171 $\pm$   10 &    2.1 $\times 10^{-14}$ &  0.43 &   7710 $\pm$    410 &    8.55 $\pm$    0.12  \\
 3\dotfill &   189 $\pm$   18 &     4.2 $\pm$  0.7 &     \nodata        &      25 $\pm$    2 &     34 $\pm$    2 &     99 $\pm$    6 &    1.1 $\times 10^{-14}$ &  0.51 &   \nodata        &     \nodata        \\   
 4\dotfill &   245 $\pm$   25 &     4.9 $\pm$  1.1 &     \nodata        &      27 $\pm$    2 &     40 $\pm$    2 &    118 $\pm$    7 &    3.8 $\times 10^{-15}$ &  0.72 &   \nodata        &     \nodata        \\   
 5\dotfill &   169 $\pm$   16 &     8.3 $\pm$  0.8 &    0.59 $\pm$ 0.08 &      28 $\pm$    2 &     64 $\pm$    4 &    194 $\pm$   11 &    1.9 $\times 10^{-14}$ &  0.67 &   8100 $\pm$    390 &    8.45 $\pm$    0.11  \\
 6\dotfill &   195 $\pm$   20 &     4.6 $\pm$  0.8 &     \nodata        &      29 $\pm$    2 &     36 $\pm$    2 &    109 $\pm$    6 &    5.4 $\times 10^{-15}$ &  0.37 &   \nodata        &     \nodata        \\   
 7\dotfill &   275 $\pm$   26 &     2.9 $\pm$  0.7 &     \nodata        &      26 $\pm$    2 &     27 $\pm$    2 &     81 $\pm$    5 &    5.7 $\times 10^{-15}$ &  0.38 &   \nodata        &     \nodata        \\   
 8\dotfill &   181 $\pm$   18 &     3.4 $\pm$  0.4 &    0.18 $\pm$ 0.04 &      26 $\pm$    2 &     37 $\pm$    2 &    112 $\pm$    7 &    2.0 $\times 10^{-14}$ &  0.29 &   6980 $\pm$    400 &    8.59 $\pm$    0.13  \\
 9\dotfill &   194 $\pm$   20 &     \nodata        &     \nodata        &      26 $\pm$    2 &     17 $\pm$    1 &     51 $\pm$    3 &    5.3 $\times 10^{-15}$ &  0.00 &   \nodata        &     \nodata        \\   
10\dotfill &   192 $\pm$   22 &     \nodata        &     \nodata        &      26 $\pm$    2 &     22 $\pm$    1 &     66 $\pm$    4 &    9.1 $\times 10^{-15}$ &  0.24 &   \nodata        &     \nodata        \\   
11\dotfill &   206 $\pm$   20 &     8.9 $\pm$  0.9 &    0.48 $\pm$ 0.12 &      25 $\pm$    2 &     59 $\pm$    3 &    177 $\pm$   10 &    3.8 $\times 10^{-14}$ &  0.77 &   7860 $\pm$    570 &    8.52 $\pm$    0.15  \\
12\dotfill &   239 $\pm$   23 &     2.5 $\pm$  0.3 &    0.30 $\pm$ 0.08 &      26 $\pm$    2 &     30 $\pm$    2 &     90 $\pm$    5 &    2.8 $\times 10^{-14}$ &  0.44 &   8300 $\pm$    660 &    8.36 $\pm$    0.16  \\
13\dotfill &   266 $\pm$   26 &     3.9 $\pm$  0.7 &     \nodata        &      26 $\pm$    2 &     27 $\pm$    2 &     81 $\pm$    5 &    9.7 $\times 10^{-15}$ &  0.22 &   \nodata        &     \nodata        \\   
14\dotfill &   197 $\pm$   19 &     3.7 $\pm$  0.6 &     \nodata        &      25 $\pm$    2 &     46 $\pm$    3 &    138 $\pm$    8 &    2.0 $\times 10^{-14}$ &  0.40 &   \nodata        &     \nodata        \\   
15\dotfill &   221 $\pm$   21 &    46.1 $\pm$  4.2 &    2.78 $\pm$ 0.32 &      25 $\pm$    2 &    145 $\pm$    9 &    444 $\pm$   26 &    6.6 $\times 10^{-15}$ &  0.19 &   9880 $\pm$    520 &    8.41 $\pm$    0.11  \\
16\dotfill &   178 $\pm$   17 &     \nodata        &     \nodata        &      25 $\pm$    2 &     17 $\pm$    1 &     52 $\pm$    3 &    9.3 $\times 10^{-15}$ &  0.00 &   \nodata        &     \nodata        \\   
17\dotfill &   240 $\pm$   23 &     6.4 $\pm$  0.8 &     \nodata        &      26 $\pm$    2 &     25 $\pm$    1 &     73 $\pm$    4 &    1.0 $\times 10^{-14}$ &  0.22 &   \nodata        &     \nodata        \\   
18\dotfill &   238 $\pm$   23 &    27.7 $\pm$  3.2 &     \nodata        &      26 $\pm$    2 &     95 $\pm$    6 &    284 $\pm$   17 &    4.0 $\times 10^{-15}$ &  0.16 &   \nodata        &     \nodata        \\   
19\dotfill &   262 $\pm$   25 &    49.4 $\pm$  4.6 &    3.31 $\pm$ 0.41 &      27 $\pm$    2 &    183 $\pm$   11 &    539 $\pm$   32 &    5.0 $\times 10^{-15}$ &  0.08 &   9810 $\pm$    530 &    8.51 $\pm$    0.11  \\
20\dotfill &   242 $\pm$   23 &     8.1 $\pm$  0.8 &     \nodata        &      26 $\pm$    2 &     45 $\pm$    3 &    134 $\pm$    8 &    2.1 $\times 10^{-14}$ &  0.41 &   \nodata        &     \nodata        \\   
21\dotfill &   243 $\pm$   23 &     3.2 $\pm$  0.4 &     \nodata        &      25 $\pm$    2 &     20 $\pm$    1 &     62 $\pm$    4 &    3.3 $\times 10^{-14}$ &  0.18 &   \nodata        &     \nodata        \\   
22\dotfill &   219 $\pm$   21 &     7.5 $\pm$  1.0 &     \nodata        &      24 $\pm$    2 &     61 $\pm$    4 &    180 $\pm$   11 &    1.5 $\times 10^{-14}$ &  0.55 &   \nodata        &     \nodata        \\   
23\dotfill &   248 $\pm$   24 &     \nodata        &     \nodata        &      26 $\pm$    2 &      1 $\pm$    0 &      4 $\pm$    0 &    1.5 $\times 10^{-15}$ &  0.16 &   \nodata        &     \nodata        \\   
24\dotfill &   224 $\pm$   23 &     \nodata        &     \nodata        &      25 $\pm$    4 &     28 $\pm$    2 &     85 $\pm$    5 &    8.7 $\times 10^{-16}$ &  1.00 &   \nodata        &     \nodata        \\   
25\dotfill &   253 $\pm$   24 &     \nodata        &     \nodata        &      25 $\pm$    2 &     12 $\pm$    1 &     34 $\pm$    2 &    3.1 $\times 10^{-15}$ &  0.15 &   \nodata        &     \nodata           
\enddata
\tablecomments{The line fluxes  are in units of H$\beta$\,=\,100. F(H$\beta$) in column (8) is the measured \hbeta\ flux, uncorrected for extinction.}
\end{deluxetable*}

%% file: tab3.tex
\begin{deluxetable}{ccc}
\tabletypesize{\scriptsize}
\tablecolumns{3}
\tablewidth{0pt}
\tablecaption{Abundances for additional H\,\scriptsize II \small regions \label{table:Subaru}}
\tablehead{
\colhead{\phantom{aaa}ID\phantom{aaa}}	     &
\colhead{\phantom{aaa}R\phantom{aaa}}	 &
\colhead{\phantom{aaa}\oh\phantom{aaa}}	 \\[0.5mm]
\colhead{}       &
\colhead{(kpc)}   &
\colhead{} \\[1mm]
\colhead{(1)}	&
\colhead{(2)}	&
\colhead{(3)}	}
\startdata
\\[-2mm]

B\,72		&		1.18		&		$8.46 \pm 0.09$	\\
B\,90		&		1.40		&		$8.47 \pm 0.08$ \\
B\,302		&		2.07		&		$8.44 \pm 0.08$

\enddata
\tablecomments{Data from the Subaru observations of \citet{Bresolin:2010}.}
\end{deluxetable}